\documentclass[a4paper, amsfonts, amssymb, amsmath, reprint, showkeys,nofootinbib,superscriptaddress,twocolumn]{revtex4-2}
\UseRawInputEncoding
\usepackage{float}
\usepackage{graphicx}
\usepackage{longtable}
\usepackage[left=15mm,right=13mm,top=23mm,columnsep=15pt]{geometry} 
\usepackage[caption=false]{subfig}
\usepackage{hyperref} 
\usepackage{multirow}
\usepackage{appendix}
\usepackage{xcolor}

\begin{document}
\title{Study of threshold anomaly in elastic scattering of $^{14}$N by $^{56}$Fe and $^{90}$Zr}
\author{K. K. Jena}
    \affiliation{P. G. Department of Physics, Bhadrak Autonomous College, Bhadrak-756100, India.}
    \affiliation{Dept. of Applied Physics and Ballistics, F. M. University, Balasore-756019, India.}

\author{S. Senapati}
    \affiliation{Dept. of Applied Physics and Ballistics, F. M. University, Balasore-756019, India.}
    
\author{B. B. Sahu}
    \affiliation{School of Applied Sciences, KIIT Deemed to be University, Bhubaneswar 751024, India.}    
    
\author{Jajati K. Nayak}
    \affiliation{Variable Energy Cyclotron Centre, 1/AF, Bidhan Nagar , Kolkata-700064, India}
    
\author{S. K. Agarwalla}
    \email[Correspondence email address: ]{san1612@rediffmail.com}
    \affiliation{Dept. of Applied Physics and Ballistics, F. M. University, Balasore-756019, India.}
\date{\today} 

\begin{abstract}
We use a phenomenological nucleus-nucleus optical potential constructed in the light of a potential developed by Ginocchio\cite{ginocchio84} to study the elastic angular distributions of different nuclear systems near Coulomb barrier. The differential cross section ratios of elastic to Rutherford are studied for systems $^{14}$N+$^{56}$Fe and $^{14}$N+$^{90}$Zr for several colliding energies. Our theoretical predictions are well competitive with the experimental data, and the significantly small imaginary part of the potential explains the elastic scattering cross section and threshold anomaly in both systems. 
\end{abstract}


\maketitle 
\section{Introduction}
Nucleus-nucleus scattering at low energy play a vital role in exploring nuclear properties. Data obtained from those scattering experiments are analyzed by various models of macro and microscopic nature. Optical model is one such established models which explain many experimental observations. The model has been extended to analyze many 
complicated nuclear phenomena \cite{hodgson86} by using the phenomenological nuclear potentials like Woods-Saxon (WS), Gaussian, modified WS, and many more. The effective potential in Optical model while explaining heavy ion scattering shows the feature of threshold anomaly near the coulomb barrier. 

To understand the elastic scattering of nuclei with optical potential, energy dependence of the  parameters are required. Threshold anomaly- an appreciable variation in real and imaginary parts of nuclear potential is observed as an important feature around the Coulomb barrier in the case of heavy-ion elastic scatterings. Many a system has been explained with rapid and localized variation of potential 
\cite{mahaux86,kobos84,nagarajan85,thompson89,stefanini87,pereira89}. The real part remains more or less stable at higher energies, but increases rapidly as the incident energy approaches the vicinity of Coulomb barrier. It attains its maximum and then decreases slowly when the incident energy falls below the barrier. Finally, the variation of the real part assumes a bell like shape around the barrier. At the same time, the imaginary part decreases from a nearly constant value around the barrier to a lower value 
\cite{nagarajan85,diaz89,fulton85,lin2001,abriola89}. The maximum value attained by the real part is just about twice the constant value it possesses at higher energies \cite{stachler91}. The change in real potential near the barrier may be due to the coupling of different elastic and quasi-elastic channels. The variation of real and imaginary parts with incident energy may also be explained through a dispersion relation as described by F. W. Byron and R. W. Fuller \cite{byron92}, which is built from the principle of causality. In this work a potential with small imaginary part (Volume part) which shows threshold anomaly is used to calculate the differential cross sections to Rutherford for various systems, as described later. 

However, when the projectile nucleus is weakly bound such as $^6$Li, the potential behavior near threshold may be different compared to tightly-bound nuclei. The imaginary part of the potential increases or remains constant with decrease in colliding energy towards the Coulomb barrier. This is widely known as Breakup Threshold Anomaly(BTA). Due to the strong coupling to the breakup channels, even at sub-barrier energies, the imaginary potential strength seems to increase as the energy is lowered down to below the barrier. The strength of real part also decreases. This has already been argued in several articles\cite{hussein2006,figueira2007,pakou2004,biswas2008,souza2007,zadro2009,fimiani2012,kumawat2008,deshmukh2011,santra2011,shaikh2016,palshetkar2014}. However, authors Rodrigo {et al.} in recent article\cite{rodrigo2019} have argued the Threshold Anomaly (TA) for weakly bound projectile system like  $^6$Li + $^{209}$ Bi and $^6$ He + $^{208}$Pb, contrary to the observations in other references. It is definitely a matter of further investigation. They have studied the problem by implementing a new method of extraction of the optical model parameters(OMP). The OMP parameters are extracted through uncertainty quantification of elastic scattering data with a physical constraint of the imaginary part. The extracted potentials can reproduce the experimental data with $\chi^2$ per degrees of freedom close to 1. They observed that the real part exhibits the usual Threshold Anomaly (TA) in $^6$Li + $^{209}$ Bi and $^6$ He + $^{208}$Pb system. Their observation of TA in above weakly-bound projectile induced reactions is contrary to other works which predicts BTA. Hence, they mention that their model can be extended to any weakly bound projectile system. It is worth to mention that these anomalies are traditional now\footnote{
From the study during early eighties of last century the indications that something unusual was happening for heavy-ion bombarding energies near the Coulomb barrier were presented by several optical model analyses while explaining elastic scattering measurements[Ref: A. Baeza, B. Bilwes, R. Bilwes, J. Diaz and IL. Ferrero, Nuci. Phys. A 419 (1984) 412; J.S. Lilley, B.R. Fulton, M.A. Nagarajan, Ii. Thompson and D.W. Banes, Phys. Lett. B 151 (1985) 181.]. The term “threshold anomaly” was applied to these observations. But these phenomena nowadays are understood and hence not “anomalous” anymore. However the anomaly term is traditionally being used like analogous “anomalous dispersion” phenomenon in optics. We also take anomaly in the same spirit.} 

However, in this article we discuss, the elastic scattering experiments of tightly bound projectile such as nitrogen, $^{14}$N with targets $^{26}$Fe and $^{90}$Zr. 

To understand near threshold behavior incase of tightly bound projectile system, several  elastic scattering experiments have been performed at various incident energies and with different target nuclei using nitrogen $^{14}$N as the projectile 
\cite{bock65,rudchik2002,towsley77,liu71,oertzen70,zeller78,yamaya88,balster87}. The results are analyzed with different theoretical approaches like optical model, double folding model {\it etc.}. However a unified theoretical approach does not exist to analyze all the experimental observations. The potentials used in some cases are associated with very large imaginary potentials, which are about 30-70\% or more of their corresponding real parts \cite{williams75}. Absorption of a major share of partial waves cannot be avoided with such potentials dealing with large imaginary part. Nevertheless, these high values of imaginary part of the potential destroy substantially the resonance states of the systems generated by the volume part of the effective potential.  


In fact Williams et al. \cite{williams75} have discussed the angular distribution from the measurements of elastic scatterings of $^{14}$N+$^{56}$Fe at energies from $E_{CM}$ = 22.4 to 32.0 MeV ($E_{Lab}$= 28 MeV to 40 MeV) and $^{14}$N+$^{90}$Zr at energies from $E_{CM}$ = 31.2 to 43.3 MeV ($E_{Lab}$= 36 MeV to 50 MeV) within the framework of optical model by considering the volume part of the imaginary potential significantly large. They achieved the best fit of experimental results of elastic scattering cross-sections of both the systems by taking imaginary potentials approximately (30-70) \% or more of real potentials. The large imaginary potential suppresses many reaction phenomena, which is clearly manifested in lighter systems. Again, this doesn't allow the optical model parameters to vary with a mass number (A) of the target nucleus \cite{williams75}. The threshold anomaly is also not explained using the above potential.

However in this article, we have used a different optical potential\cite{sahu2003,mallick2006} with reasonable imaginary part based on Ginocchio potential \cite{ginocchio84}. This has an analytically solvable asymmetric part which is more versatile and can be applied for various types of projectiles. We explain the differential cross-section ratios of elastic to Rutherford ($\sigma_{el}/\sigma_{Ruth}$) for $^{14}$N+$^{56}$Fe and $^{14}$N+$^{90}$Zr at various energies. One of the present authors have also applied this potential \cite{sahu2003,mallick2006} to study $^{16}$O+$^{28}$Si and $^{12}$C+$^{24}$Mg systems. To check the wide applicability of the potential, we, in this regard, choose heavy targets like Ferrum, $^{56}$Fe and Zirconium,  $^{90}$Zr with nitrogen $^{14}$N as the projectile. This is because the experimental observations are available for these targets so that theoretical calculations can be compared. Our theoretical results with potential\cite{sahu2003,mallick2006} having small imaginary part explain the data of elastic scatterings at different energies. We also observe the phenomenon of threshold anomaly in these systems. The threshold anomaly phenomena are not shown explicitly for the systems $^{14}$N+$^{56}$Fe and $^{14}$N+$^{90}$Zr in the referred literature \cite{williams75}.

In section II, we present the formulation of our potential. In Section III we discuss the results from our calculation and compare with the elastic scattering data of $^{14}$N with $^{26}$Fe and $^{90}$Zr. We present the summary and conclusions in Section IV.

\section{THEORETICAL FORMULATION}
The interactions between nuclei can be described by using a potential that consists of Coulomb potential and nuclear potential. Taking centrifugal force into consideration, the effective potential in a nucleus-nucleus 
collision can be written as,	 
$$V_{eff} = V_C(r) + V_N (r) + V_{CF}$$
where,	$V_C(r)$= potential due to Coulomb force between two nuclei, 
$V_N(r)$= potential due to nuclear force. $V_{CF}$= potential due to 
centrifugal force and which can be described as follows; 
$$V_{CF} \sim \frac{l(l+1){\hbar}^2}{2\mu r^2}$$ where $l$ is the orbital 
quantum number and $\mu$ is the reduced mass of the projectile and 
target, i.e., $m_1 m_2/(m_1+m_2)$.

Thus effective potential with reduced mass μ for a nucleus-nucleus 
scattering is given by 
\begin{equation}
V_{eff} =V_C(r)+V_N(r)+\frac{l(l+1){\hbar}^2}{2\mu r^2}
\label{eq_effectivepot}
\end{equation}

Nuclear part of the optical potential in the complex form is given by, 
$V_N(r) = V_n(r) + iW_n(r)$. The real part $V_n(r)$ of the potential 
assumes the form \cite{sahu2003,mallick2006},

\begin{equation}
V_n(r)= \left\{ \begin{array}{ll}
           \frac{-V_B}{B_1}\left[B_0+(B_1-B_0)(1-y_1^2) \right]  &  \text{if}~~ 0<r<R_0\\
           & \\
           \frac{-V_B}{B_2} \left[B_2(1-y_2^2)\right] & \text{if}~~r\ge R_0\\
          \end{array}
 \right.
 \label{eq_nuclearpot}
\end{equation}

with substitutions of $y=tanh\rho_n$, $\rho_n=(r-R_0)b_n$, we get
\begin{equation}
V_n(r)= \left\{ \begin{array}{ll}
           -\frac{V_B}{B_1}\left[B_0+\frac{(B_1-B_0)}{cosh^2\rho_1} \right]  &  \text{if}~~ 0<r<R_0\\
           & \\
           -\frac{V_B}{B_2}\left[\frac{B_2}{cosh^2\rho_2} \right] & \text{if}~~r\ge R_0\\
          \end{array}
 \right.
 \label{eq_nuclearpot2}
\end{equation}

\begin{figure}
    \centering
    \begin{minipage}{0.45\textwidth}
        \centering
        \includegraphics[width=0.9\textwidth]{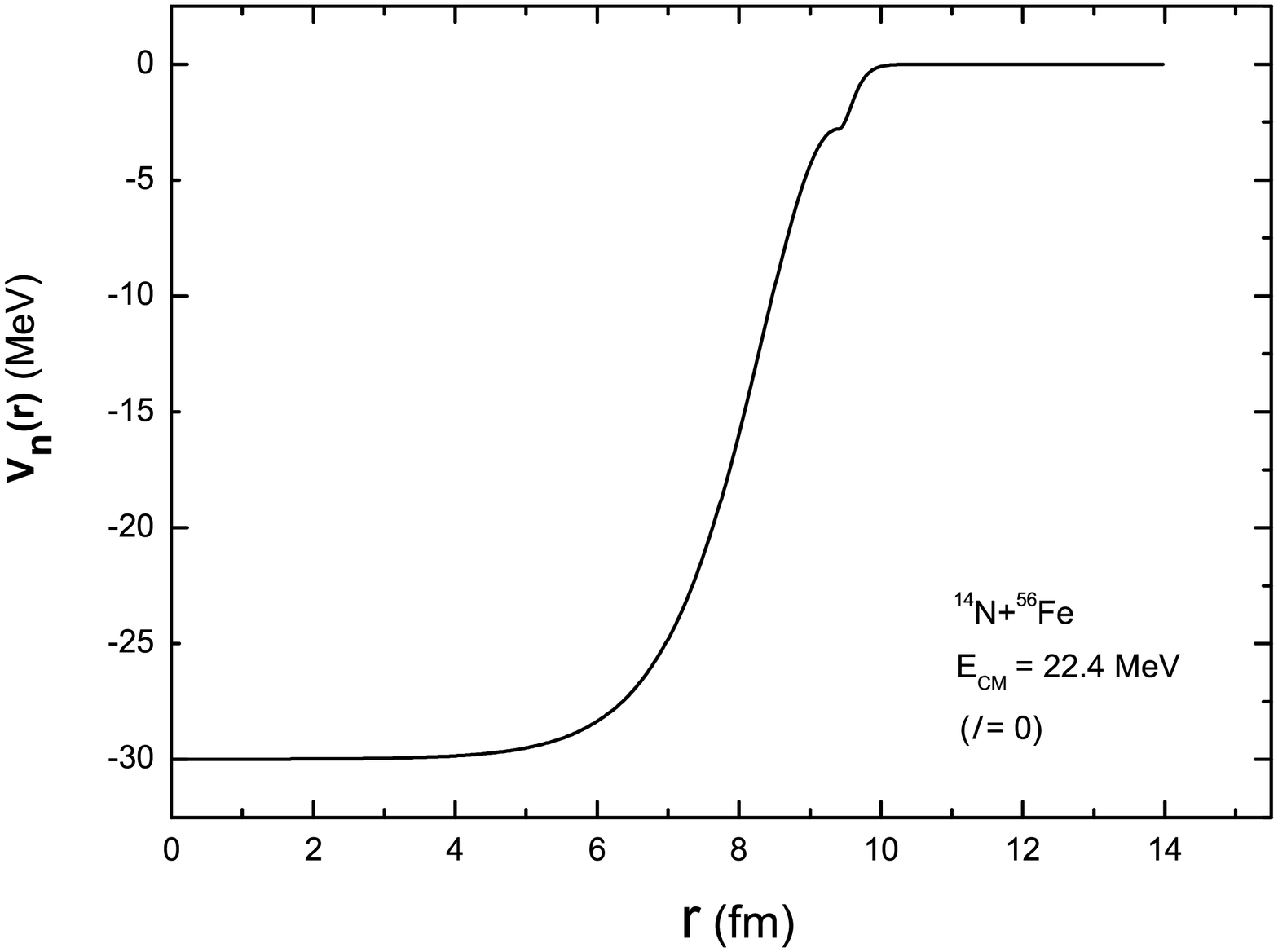} 
        \caption{Plot of real part $V_n(r)$ of optical potential for the 
	    system $^{14}$N + $^{56}$Fe as described in 
	    Eq.\ref{eq_nuclearpot2} for energy $E_{CM} = 22.4$ MeV. 
	    Values of parameters are set at $R_0=9.4$ fm, $B_1=4.0$, 
	    $B_2 = 0.01, B_0 = 29.99$ MeV and $V_B$= 2.8 MeV}
        \label{fig1}
    \end{minipage}\hfill
    \begin{minipage}{0.45\textwidth}
        \centering
        \includegraphics[width=0.9\textwidth]{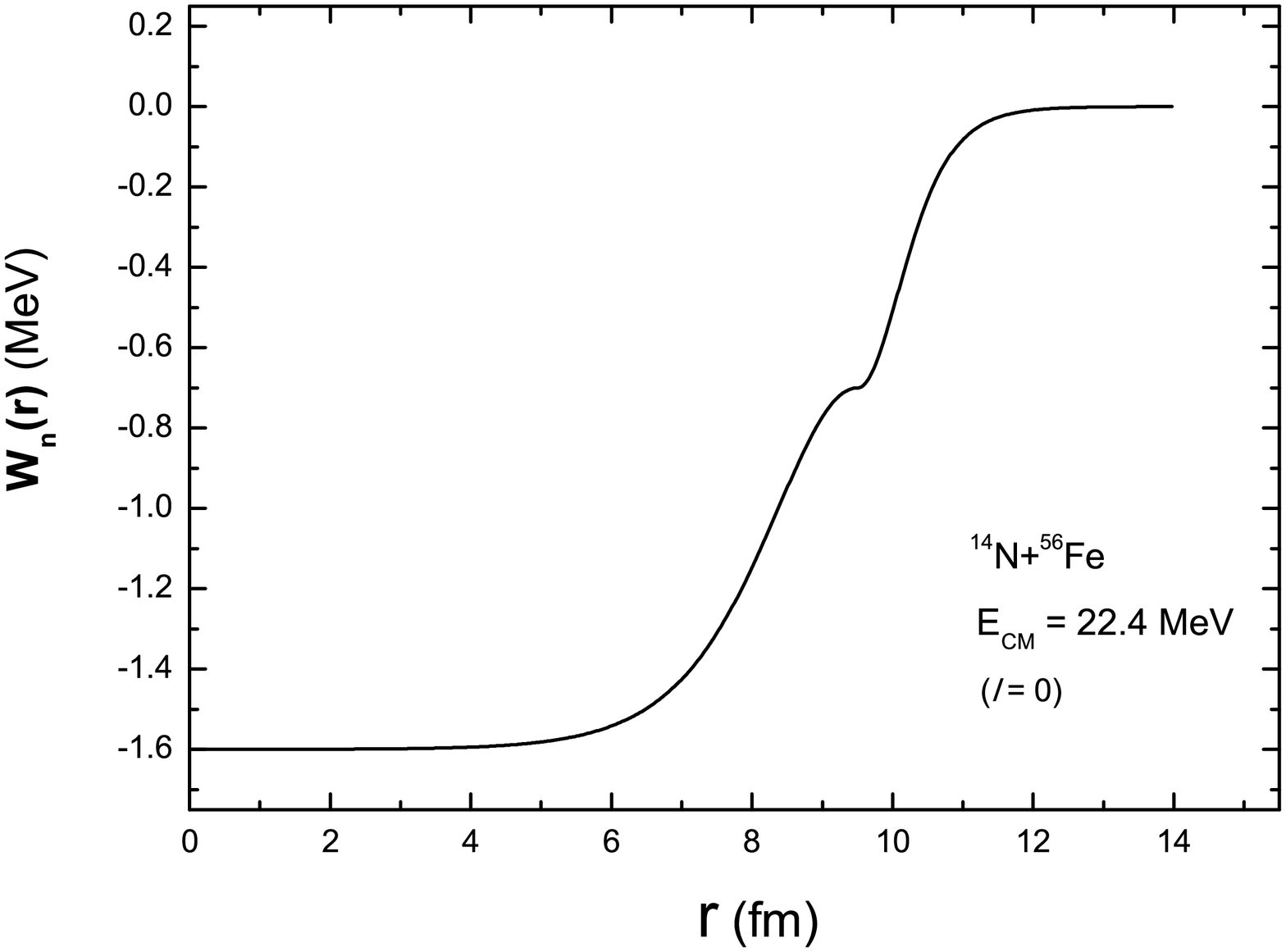} 
        \caption{Plot of imaginary part $W_n(r)$ of the optical potential 
	    for the system $^{14}$N+$^{56}$Fe for energy $E_{CM}$ = 22.4 MeV. 
	    Values of the parameters are $R_{0W}$ = 9.5 fm, $W_1$=1.1, 
	    $W_2$ = 0.28, $W_0$ = 1.6 MeV and $V_{BW}$ = 0.7 MeV.}
        \label{fig2}
    \end{minipage}
\end{figure}

The slope parameters are given by $b_n=\frac{\sqrt{2mV_B}}{\hbar^2B_n}$ ($n = 1, 2$). The parameter $R_0$ is the radial distance in the surface region which is close to the position of the effective S-wave barrier potential described in Eq.\ref{eq_effectivepot}. $B_0$ and $V_B$ control the depth of the potential at $r = 0$ and $R_0$ respectively. The parameters $B_n$ and $V_B$ control the slope parameter $b_n$ on either side of $R_0$. Thus, a small value of $B_n$ allows the potential to have sharp variation, when $b_n$ undergoes a large change. The real part of the potential $V_n(r)$ with a set of parameters is shown in Fig.\ref{fig1}.  

The nuclear potential decreases monotonically with $r$ in the case of standard WS form. This potential shows a slight non-trivial behavior near $r = R_0$, where the two parts of the potential corresponding to the interior region with slope  $b_1$ and the outer region with different slope  $b_2$  are connected together to satisfy analytic continuity at $r = R_0$. This consideration ensures continuity in two 
parts of the potential at that position. Such a nuclear potential takes care of the various phenomena, namely, the resonance phenomena belonging to the formation of the composite binuclear system, effects of frictional forces, 
and transfer of one or cluster of nucleons from the target to the projectile and/or vice versa in this configuration, when the two bombarding nuclei touch each other in the surface region around $r = R_0$. The necessity of non-trivial behavior of our potential around  $r = R_0$ was realised so as to explain wide range of experimental data of differential scattering cross-sections. Such non-trivial behaviour explains the data over a wide range of energies for the two systems without creating any irregularities in the variation of amplitude of wave function with radial distance. 

\begin{figure}
    \centering
    \begin{minipage}{0.45\textwidth}
        \centering
        \includegraphics[width=0.9\textwidth]{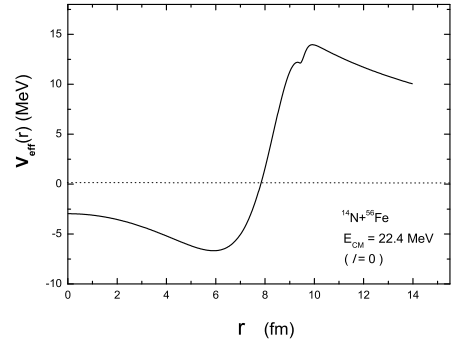} 
        \caption{Plot of effective optical potential as described in 
	    Eq.\ref{eq_effectivepot} for $l=0$ for $^{14}$N+ $^{56}$Fe 
	    at energy $E_{CM}$ = 22.4 MeV. Set of values of the parameters 
	    are $R_0$=9.4 fm, $B_1$=4.0, $B_2$ = 0.01, $B_0$ = 29.99 MeV, 
	    $V_B$= 2.8 MeV, $R_{0W}$ = 9.5 fm, $W_1$=1.1, $W_2$ = 0.28, 
	    $W_0$ = 1.6 MeV and $V_{BW}$ = 0.7 MeV.}
        \label{fig3}
    \end{minipage}
\end{figure}

The imaginary part $W_n(r)$  of the present optical potential has the same form as that of the real part but with different strengths. 
The imaginary part can be expressed by the following Eq.\ref{eq_nuclearpot3}. 
\begin{equation}
W_n(r)= \left\{ \begin{array}{ll}
           -\frac{V_{BW}}{W_1}\left[W_0+\frac{(W_1-W_0)}{cosh^2\rho_1} \right]  &  \text{if}~~ 0<r<R_{0W}\\
           & \\
           -\frac{V_{BW}}{W_2}\left[\frac{W_2}{cosh^2\rho_2} \right] & \text{if}~~r\ge R_{0W}\\
          \end{array}
 \right.
 \label{eq_nuclearpot3}
\end{equation}
The imaginary part $W_n(r)$ of the potential is plotted in Fig.\ref{fig2} 
with a suitable set of parameters. The Coulomb potential between the 
two colliding nuclei is given as

\begin{equation}
V_C(r)= \left\{ \begin{array}{ll}
           \frac{Z_p Z_T e^2}{2R_C^3}(3R_C^2-r^2)  &  \text{if}~~ r<R_C\\
           & \\
           \frac{Z_p Z_T e^2}{r} & \text{if}~~r > R_C\\
          \end{array}
 \right.
 \label{eq_coulombpot}
\end{equation}
where, 	$R_C = r_C ( {A_P}^{1/3} + {A_T}^{1/3})$; $Z_P$ and $Z_T$ are the 
atomic numbers of projectile and target nuclei respectively; $A_P$ and 
$A_T$ are the mass numbers of projectile and target nuclei respectively. 
The value of $r_C$ is taken to be 1.25 fm. With $V_N(r)$, and $V_C(r)$, 
the real part of the effective potential $V_{eff}(r)$ for $l=0$ is 
plotted in Fig.\ref{fig3} with the same set of parameters considered 
for the real and imaginary parts.

We solve the Schrodinger equation to obtain the total scattering 
amplitude $f(\theta)$.

\begin{equation}
 \left[\frac{-\hbar^2}{2\mu}\nabla^2 + V_{eff}(r) \right]\psi( \vec r)=E \psi(\vec r)
 \label{eq_schrodinger}
\end{equation}

Total scattering amplitude $f(\theta$) is expressed as the sum of Coulomb scattering amplitude $f_C(\theta)$ and nuclear scattering amplitude 
$f_N(\theta)$. Thus,
\begin{equation}
 f(\theta)=f_C(\theta) + f_N(\theta)
\label{eq_scat}
\end{equation}
Where the nuclear amplitude $f_N(\theta)$ has the expansion

\begin{equation}
 f_N(\theta)=\frac{1}{2ik}\sum_i(2l+1)e^{2i\sigma_i}\left(e^{2i\bar{\delta}-1} \right)P_l(cos\theta)
\label{eq_scat2}
\end{equation}
Here $\sigma_i$ is the Coulomb phase shift due to scattering and 
$\bar{\delta_i}$ is the nuclear phase shift. The measured differential 
elastic scattering cross-sections ratio to Rutherford is given by
\begin{equation}
\frac{d\sigma}{d\sigma_R}=|{\frac{f(\theta)}{f_C(\theta)}}|^2
\label{eq_scat3}
\end{equation}
The elastic scattering cross-section $\sigma_{el}$ and reaction 
cross-section $\sigma_{rl}$ for $l^{th}$ partial wave are given as
\begin{equation}
\sigma_{el}=\frac{\pi}{k^2}(2l+1)|1-S_l|^2
\label{eq_sigmael}
\end{equation}
\begin{equation}
\sigma_{rl}=\frac{\pi}{k^2}(2l+1)(1-|S_l|^2)
\label{eq_sigmarl}
\end{equation}
Where $S_l$ is the S-matrix for the $l$th partial wave. With above theoretical formalism and potential, the elastic scattering results are discussed below.
\section{RESULTS}
In \cite{williams75,aygun17}, authors have performed the analysis of elastic scattering data of $^{14}$N + $^{56}$Fe and $^{14}$N + $^{90}$Zr at various energies. Williams {\it et al.} \cite{williams75} have taken Wood Saxon based optical potential where the imaginary potential depths are approximately 30-60\% or more that of the real part, whereas M. Aygun\cite{aygun17} has studied many systems with $^{14}N$ target including this $^{14}$N + $^{90}$Zr using microscopic double folding model and extracted new equation which gives the imaginary potential that explains scattering data of different systems. The imaginary potential is evaluated by double folding potential and contains two normalisation constants($N_i, N_r$). This is like taking two more free parameters into the calculation. 


We apply the formalism with optical potential discussed in Section-II for the analysis of angular cross-section of elastic scattering of $^{14}$N + $^{56}$Fe and $^{14}$N + $^{90}$Zr systems at their respective energies of incidence. The variations of real and imaginary parts of the potential are investigated about the Coulomb barriers of all these systems.
\subsection{Elastic scattering results}
\subsubsection{System $^{14}$N + $^{56}$Fe}
\begin{figure}
    \centering
    \begin{minipage}{0.45\textwidth}
        \centering
        \includegraphics[width=0.9\textwidth]{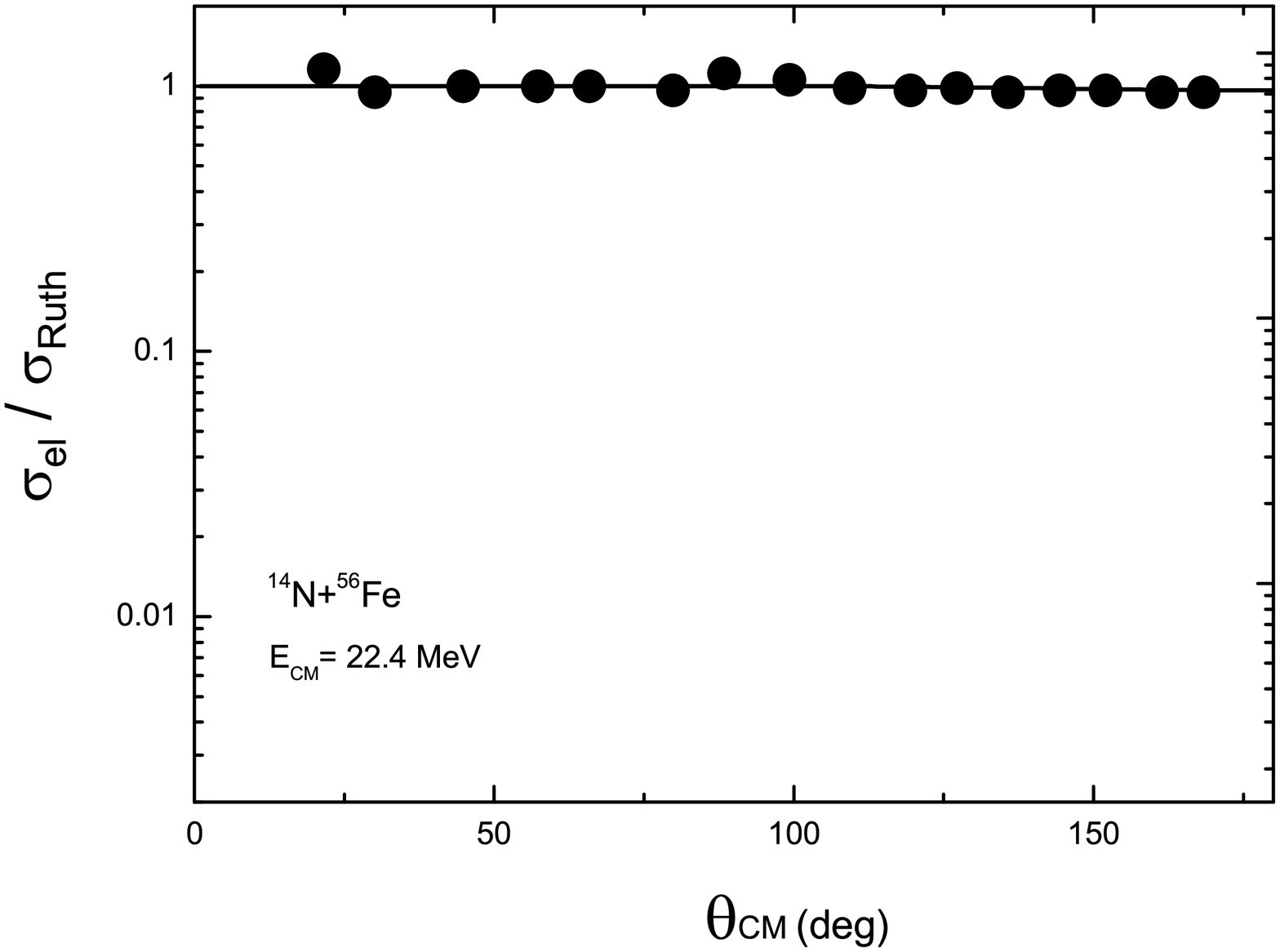} 
        \includegraphics[width=0.9\textwidth]{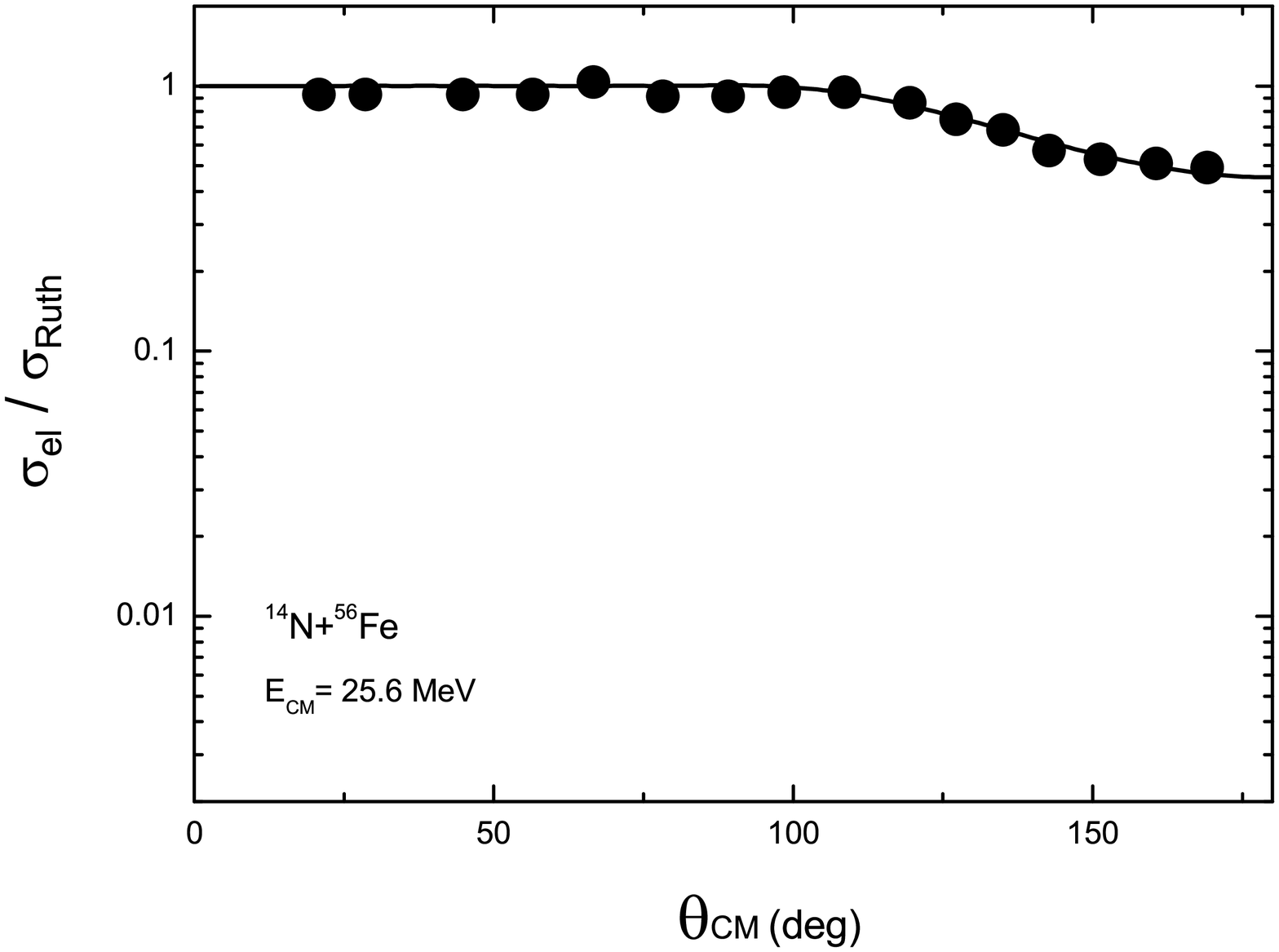}
        \includegraphics[width=0.9\textwidth]{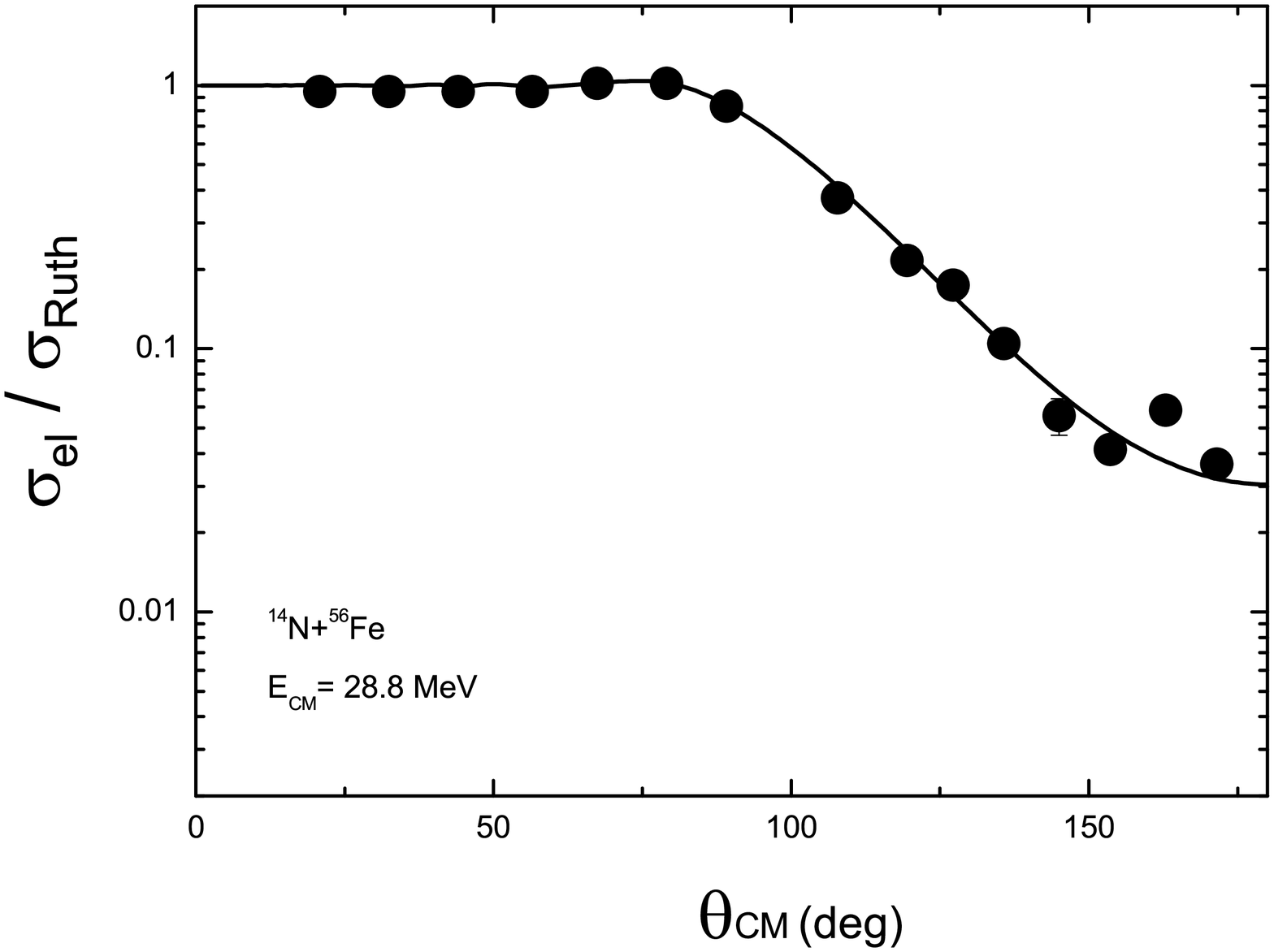}
        \includegraphics[width=0.9\textwidth]{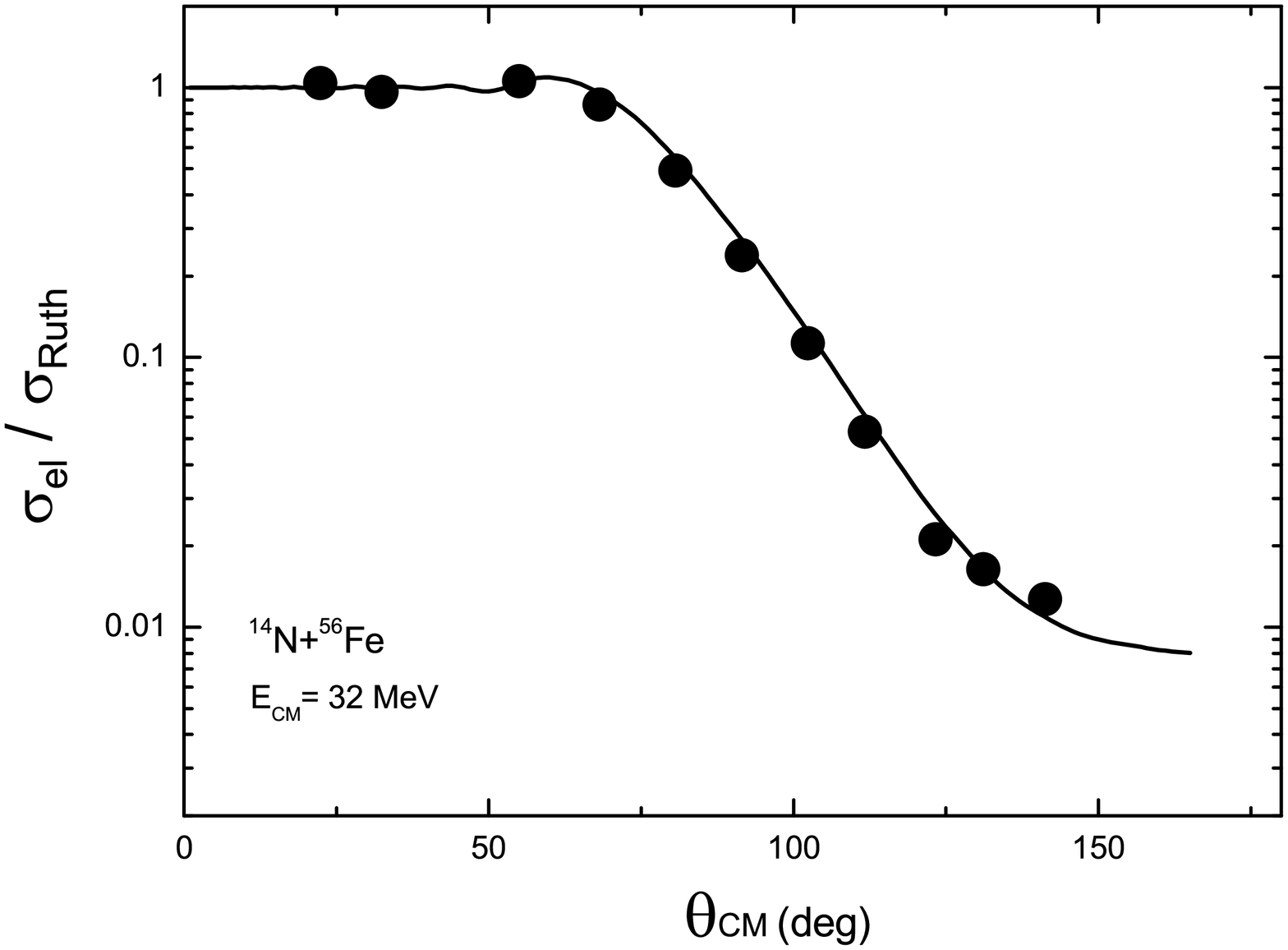}
        \caption{The comparison of calculated angular elastic scattering 
	    cross-sections with experimentally measured ones at four 
	    incident energies of $^{14}$N+$^{56}$Fe. The solid circles 
	    represent experimental values, whereas, the solid line curves 
	    represent theoretically calculated values.The experimental values 
	    are taken from ${http://nrv.jinr.ru}$. Values of independent 
	    parameters are taken to be $R_0$=9.4 fm, $R_{0W}$=9.5 fm, 
	    $B_1$=4.0, $W_0$=1.6 MeV, $W_1$=1.1 and $W_2$=0.28.}
        \label{fig4}
    \end{minipage}
\end{figure}

Energy range for elastic scattering of  $^{14}$N ions by $^{56}$Fe target are taken from 28 to 40 MeV in laboratory frame, i.e., 22.4 to 32 MeV in center of mass frame, as the scattering system has its Coulomb barrier at about 25 MeV ($E_{CM}$). The Coulomb barrier is evaluated using Bass formula \cite{bass74}. To explain the experimental observation of angular cross section the parameters are tuned to obtain the best-fit of the optical potential and are given in Table-\ref{table_paramet_fe}. Colliding energy in center of mass frame($E_{CM}$), potential parameters$V_B$ , $B_0$ and $V_{BW}$ have the same unit in MeV. 
The results of angular elastic scattering cross-sections for this system are compared with the experimentally measured values\cite{williams75}({\it http://nrv.jinr.ru}).
\begin{table}
\caption{Energy dependent parameters for $^{14}$N+$^{56}$Fe system}
 \begin{tabular}{ |c|c|c|c|c| }
  \hline
    $E_{CM}$ & $V_B$ & $B_0$  & $B_2$ & $V_{BW}$ \\
  (MeV)& (MeV) & (MeV) & (MeV) & (MeV) \\
  \hline
  22.4 & 2.8 & 29.99 & 0.01 & 0.7 \\
  \hline
   25.6 & 3.8 & 26.99 & 0.15 & 1.2 \\
  \hline
  28.8 & 2.4 & 22.49 & 0.17 &1.4 \\
  \hline
  32.0 & 2.1 & 20.99 & 0.095 & 1.4 \\
  \hline    
 \end{tabular}
 \label{table_paramet_fe} 
\end{table}
Six out of ten parameters remain energy-independent, but $V_B, B_0, B_2, V_{BW}$ change with energy. The values of the independent parameters are found to be $R_0$=9.4 fm, $R_{0W}$=9.5 fm, $B_1$=4.0, $W_0$=1.6 MeV, $W_1$=1.1 and $W_2$=0.28 while finding the best-fit with the data for the entire range of energies. The other four parameters mentioned in the above Table-\ref{table_paramet_fe} vary with incident energies. The theoretical results of ${\sigma_{el}}/{\sigma_{Ruth}}$ with $\theta_{CM}$ are shown in solid line. As depicted in the Fig.\ref{fig4}, present calculation  agrees well with the experimental observations.
In addition to the fact that our optical potential takes care of both volume and surface region, the best fit imaginary potentials are very small as compared to the real potentials. The ratio of imaginary to real part is also given in Table-\ref{table_potential_fe}. The ratio remains 
below 8\%  which is much less compared to the calculation in Ref.\cite{williams75}(30-80\%). As we know that higher value in the ratio of imaginary to real part may suppresses some resonance states of the system generated by the effective potential.
\begin{table}
\caption{Ratio between imaginary to real part of the potential}
 \begin{tabular}{ |c|c|c|c|c| }
  \hline
  $E_{CM}$ & Real part & Imaginary part & Ratio & Ratio in Ref.\cite{williams75} \\
    & & & at r=0, (in \%) & (in \%) \\
  \hline
  22.4 & 29.99 & 1.6 & 5.3 & 30 \\
  \hline
  25.6 & 26.99 & 1.6 & 5.9 & 32.96 \\
  \hline
  28.8 & 22.49 & 1.6 & 7.1 & 53.77 \\
  \hline
  32.0 & 20.99 & 1.6 & 7.6 & 78.57 \\
 \hline    
 \end{tabular}
 \label{table_potential_fe} 
\end{table}
\subsubsection{System $^{14}$N+$^{90}$Zr}
\begin{figure}
    \centering
    \begin{minipage}{0.45\textwidth}
        \centering
        \includegraphics[width=0.8\textwidth]{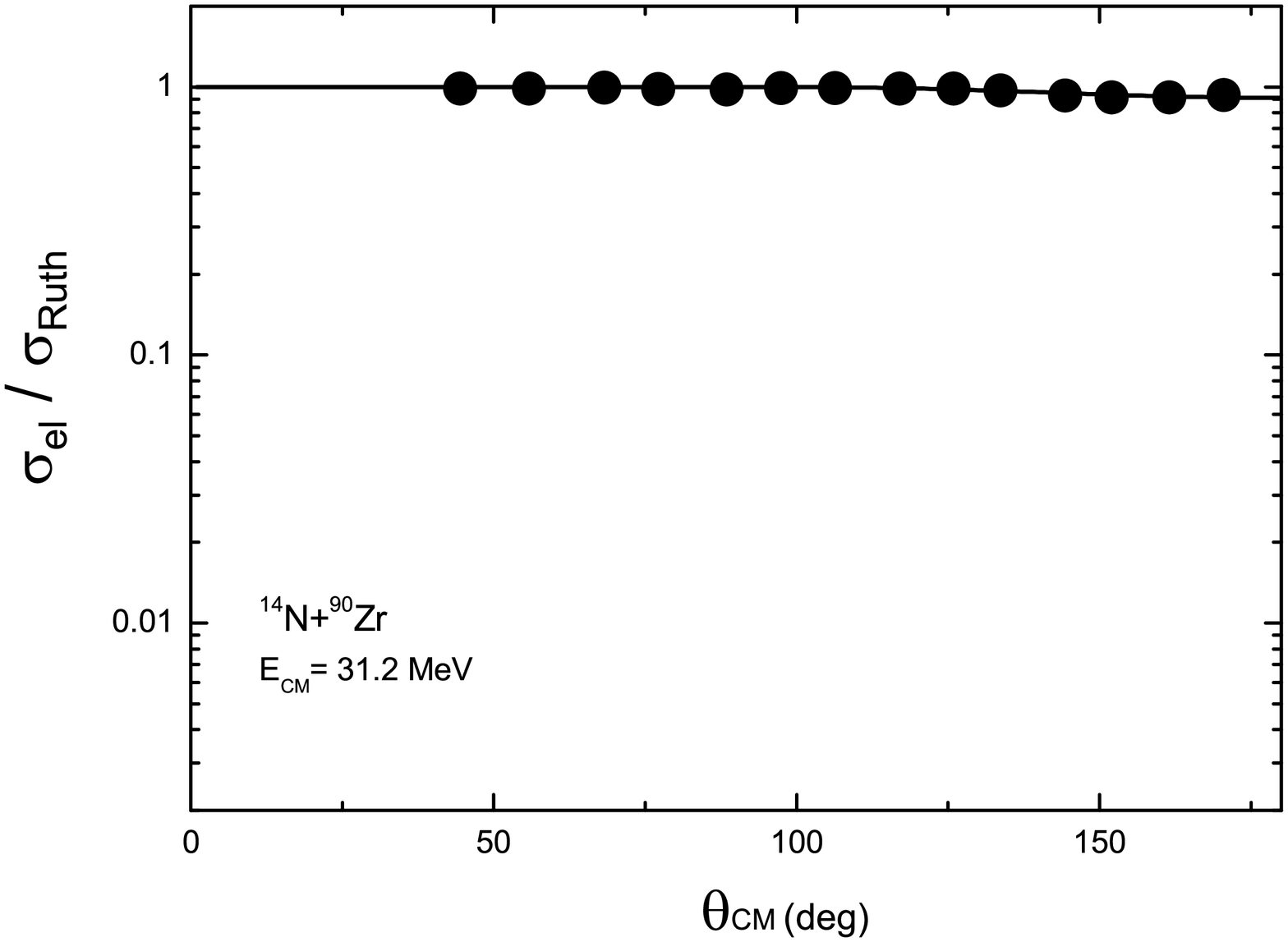} 
        \includegraphics[width=0.8\textwidth]{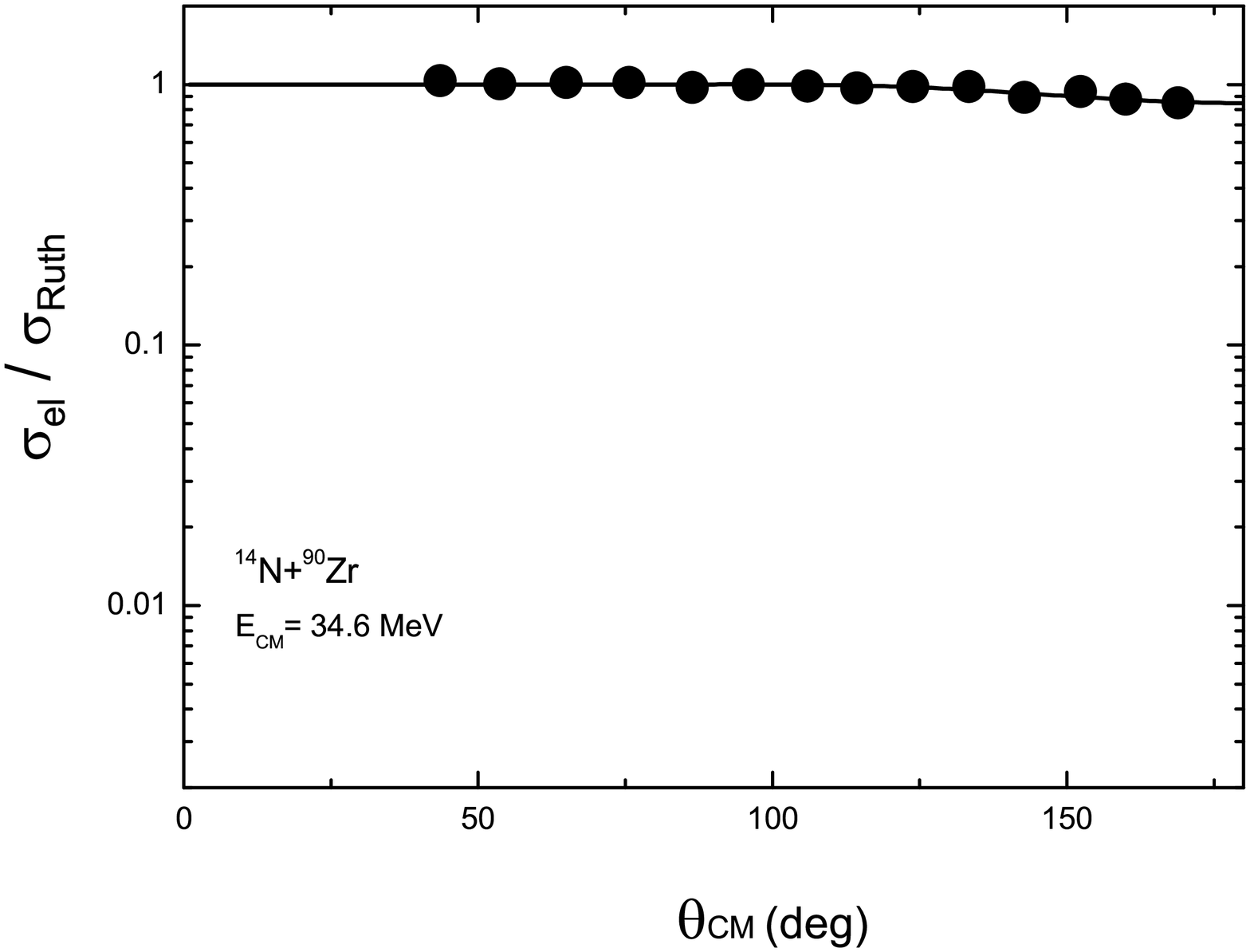}
        \includegraphics[width=0.8\textwidth]{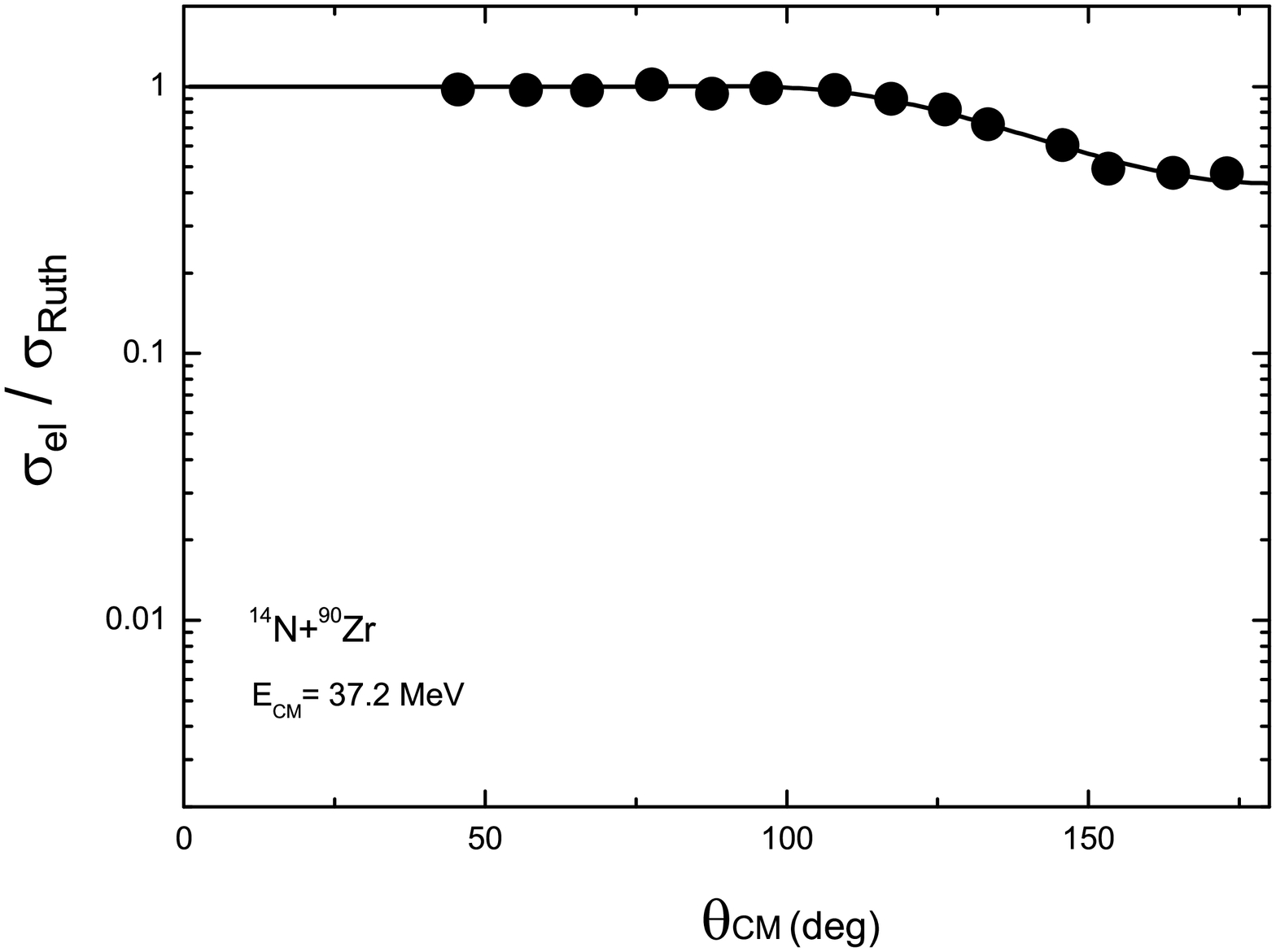}
        \includegraphics[width=0.8\textwidth]{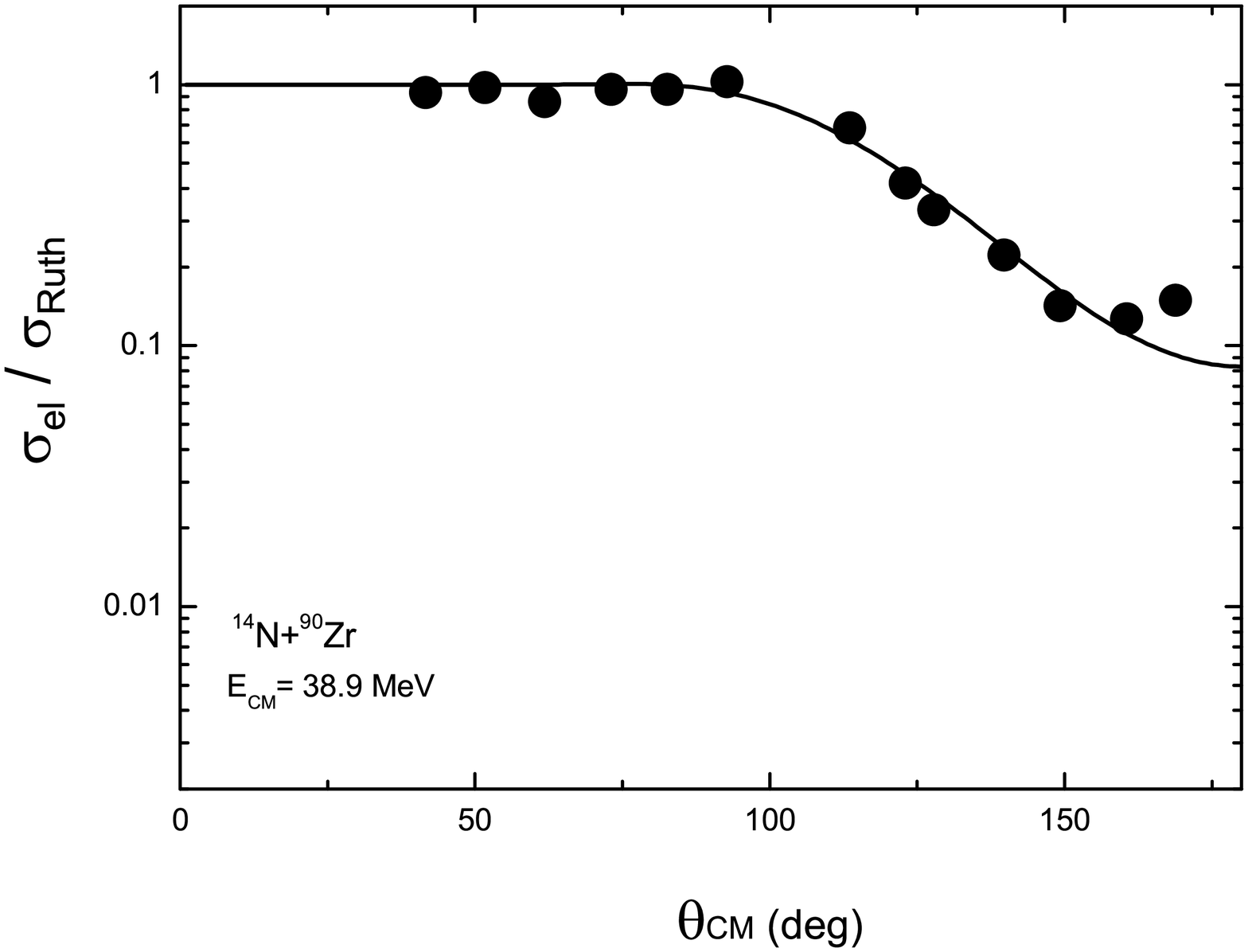}
        \includegraphics[width=0.8\textwidth]{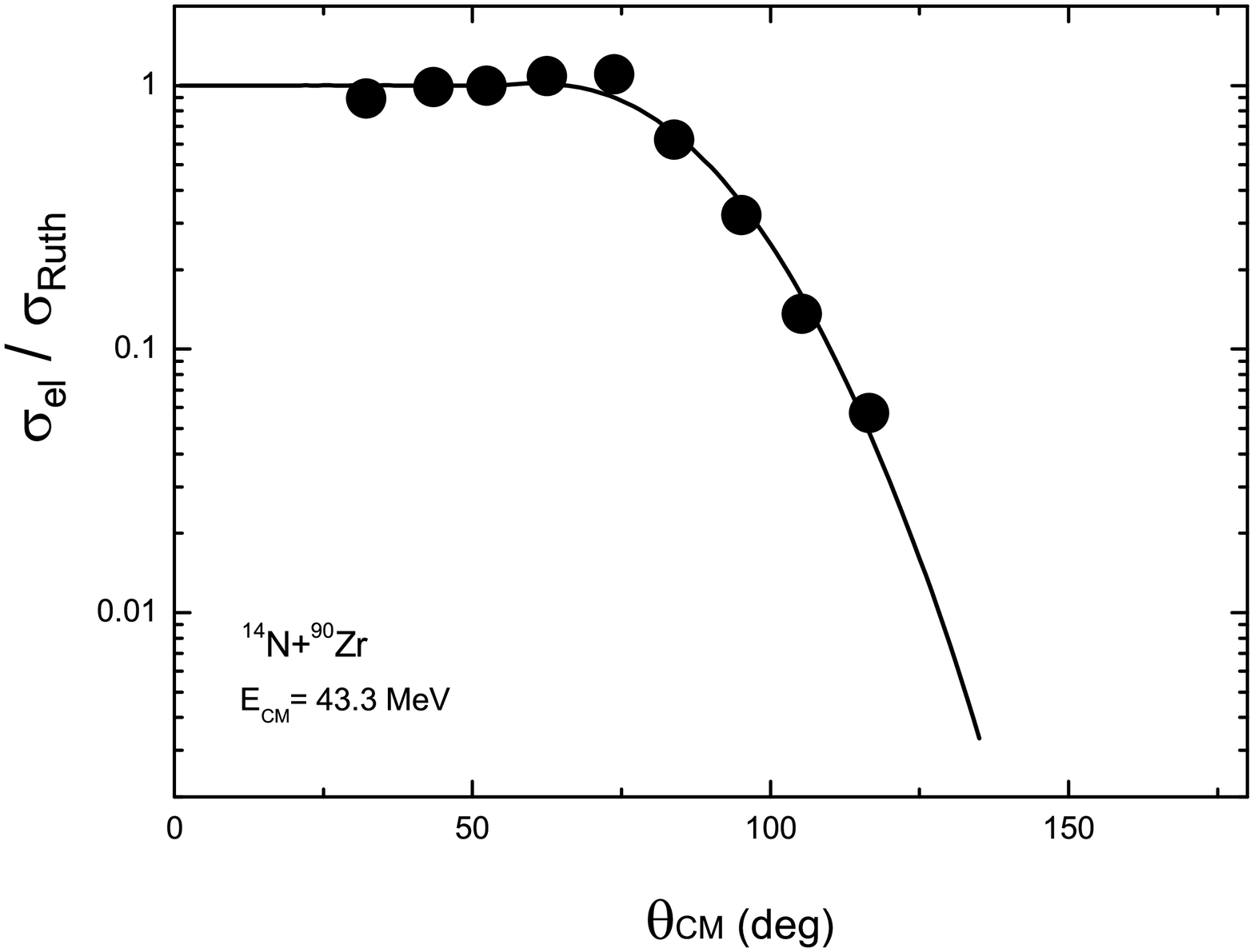}
        \caption{Comparison of theoretically calculated angular elastic 
	    scattering cross-sections with experimentally measured ones 
	    at five incident energies of the system $^{14}$N+$^{90}$Zr. 
	    The small solid circles represent experimental values and 
	    the solid line curves represent our calculated values. 
	    The experimental values are taken from ${http://nrv.jinr.ru}$.
	    Values of independent parameters are taken to be $R_0$=9.9 fm, 
	    $R_{0W}$=10.7 fm, $B_1$=4.0, $B_2$=0.11, $W_0$=2.2 MeV and 
	    $W_1$=1.0.}
        \label{fig5}
    \end{minipage}
\end{figure}

In a similar way to the analysis of the above system, energy range for the scattering of  $^{14}$N ions by $^{90}$Zr target is taken from 36 to 50 MeV in laboratory frame. In case of center of mass frame this is from 31.2 to 43.3 MeV. The scattering system has its Coulomb barrier at about 36 MeV ($E_{CM}$) using Bass formula\cite{bass74}. 
The best-fit parameters of the potential in this system are presented in Table-\ref{table_paramet_zr}. The angular cross-sections are compared with experimentally measured data \cite{williams75}(${http://nrv.jinr.ru}$). 
\begin{table}
\caption{Energy dependent parameters for $^{14}$N+$^{90}$Zr system}
 \begin{tabular}{ |c|c|c|c|c| }
  \hline
    $E_{CM}$ & $V_B$ & $B_0$  & $W_2$ & $V_{BW}$ \\
  (MeV)& (MeV) & (MeV) & (MeV) & (MeV) \\
  \hline
  31.2 & 2.46 & 25.0 & 0.60 & 0.6 \\
  \hline
  34.6 & 4.0 & 22.2 & 0.17 & 1.2 \\
  \hline
  37.2 & 3.8 & 27.7 & 0.13 & 1.5 \\
  \hline
  38.9 & 2.5 & 28.4 & 0.31 & 1.45 \\
  \hline
  43.3 & 2.2 & 29.8 & 0.45 & 1.5 \\
  \hline    
 \end{tabular}
 \label{table_paramet_zr} 
\end{table}
Parameters, $R_0$=9.9 fm, $R_{0W}$=10.7 fm, $B_1$=4.0, $B_2$=0.11, $W_0$=2.2 MeV and $W_1$=1.0 like the previous analysis are also found to be energy independent when the results are compared with measurments. The other four parameters mentioned in Table-\ref{table_paramet_zr} vary 
with incident energies. Those are $W_2$, $V_B$ , $B_0$ and $V_{BW}$. Coulomb radius is taken to be $r_C$ = 1.25 fm. The theoretical results are then plotted in Fig.\ref{fig5}. Solid line represents theoretical calculation and solid circles are for experimental data. The agreement is fairly well.
Besides the fact that present optical potential takes care of both volume and surface region, the imaginary potentials used in the best fit to the data are very small in comparison to their real part. The ratio of imaginary part to real part is given in Table-\ref{table_potential_zr}. The ratio remains below 10\% of the real potential within the volume region in our calculations where the same ratio varies from 30 to 175\% in Ref.\cite{williams75}.
\begin{table}
\caption{Ratio between imaginary to real part of the potential for 
	$^{14}$N + $^{90}$Zr system.}
 \begin{tabular}{ |c|c|c|c|c| }
  \hline
  $E_{CM}$ & Real part & Imaginary part & Ratio & Ratio (in \%) \\
    & & & at r=0, (in \%) & in Ref.\cite{williams75} \\
  \hline
  31.2 & 25.0 & 2.2 & 8.8 & 175.2 \\
  \hline
  34.6 & 22.2 & 2.2 & 9.9 & 38.28 \\
  \hline
  37.2 & 27.7 & 2.2 & 7.9 & 30.68  \\
  \hline
  38.9 & 28.4 & 2.2 & 7.7 & 57.04 \\
 \hline    
 43.3 & 29.8 & 2.2 & 7.3 & 44.63 \\
 \hline
 \end{tabular}
 \label{table_potential_zr} 
\end{table}
\subsection{Phenomena of Threshold Anomaly}
\begin{figure}
    \centering
    \begin{minipage}{0.45\textwidth}
        \centering
        \includegraphics[width=0.9\textwidth]{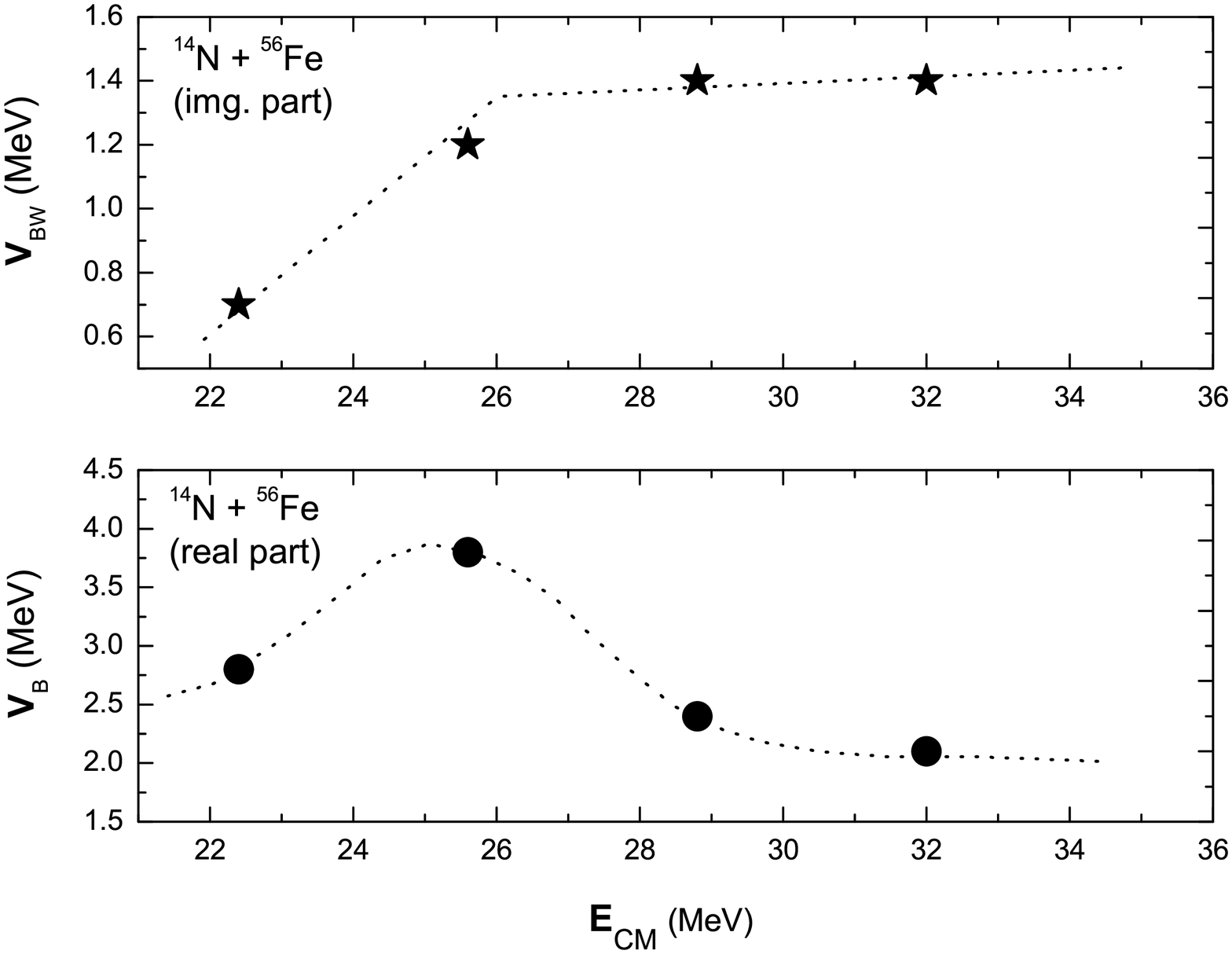} 
        \includegraphics[width=0.9\textwidth]{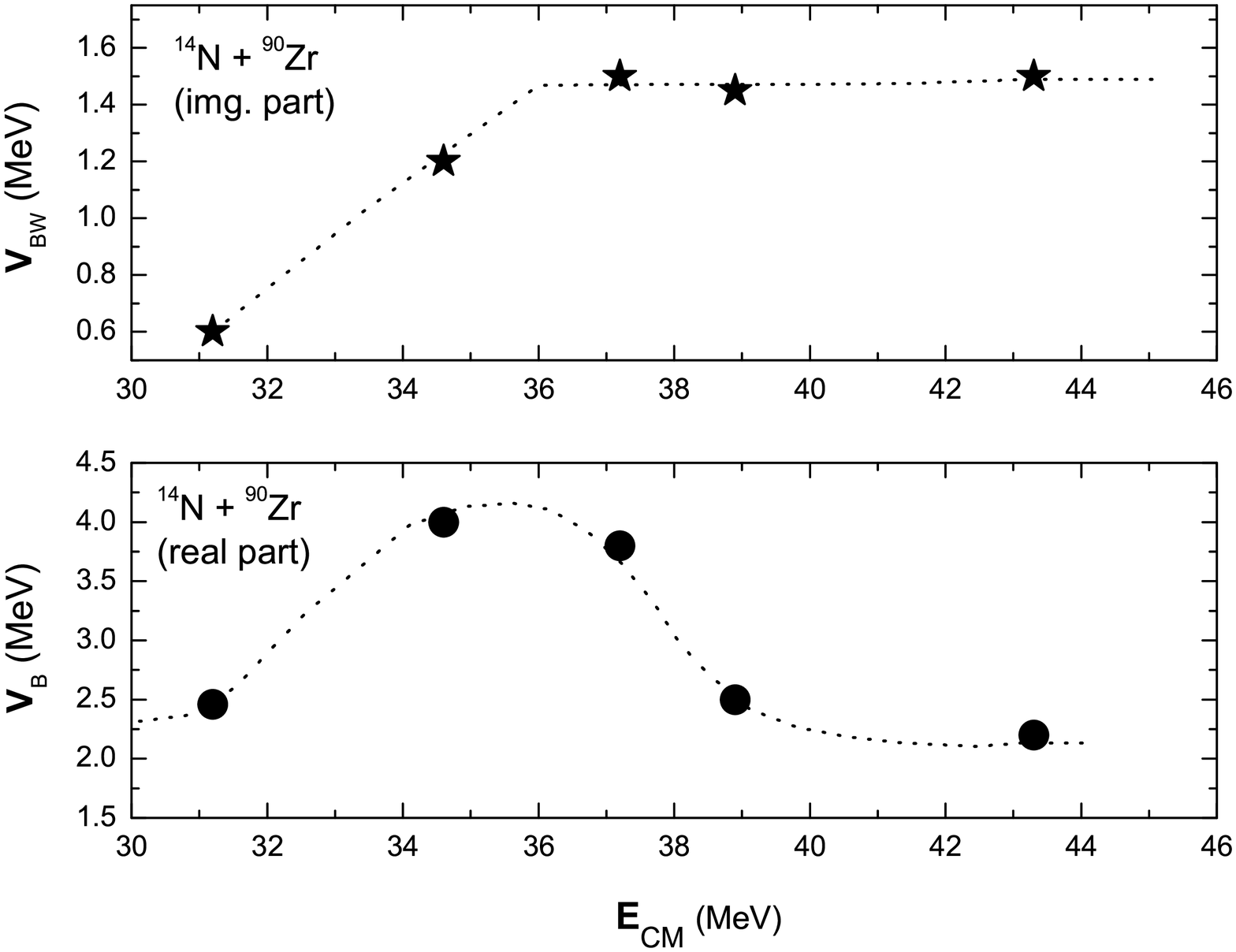}
        \caption{The variation of real part ($V_B$) and imaginary part ($V_{BW}$) of the optical potential around the Coulomb  	    barriers of the two systems are shown. The dotted-line with small star-symbols represents the variation of imaginary parts 
	    with energy, whereas, the dotted-line with solid circle-symbols 
	    represents the simultaneous variation of real parts.}
        \label{fig6}
    \end{minipage}
\end{figure}

While comparing the theoretical calculation with experimental observations for several colliding energies, we find variations in real and imaginary parts of the potential near Coulomb barrier. The variation is almost similar in both the systems. Such variations are shown in Fig.\ref{fig6}. This is nothing but the threshold anomaly which has been argued for tight nuclei projectile system like $^{14}$N in several articles. The same anomaly feature is exhibitted by the present potential. 
The presence of threshold anomaly in the elastic scattering of the systems $^{14}$N+$^{56}$Fe and $^{14}$N+$^{90}$Zr are not explicitly shown in reference \cite{williams75}. But our calculation in the case of $^{14}$N+$^{56}$Fe finds the rise in magnitude of imaginary part ($V_{BW}$) of the optical potential with incident energy till the peak of Coulomb barrier and then shows a saturation at higher energies. The real part changes rapidly from 2.8 MeV to 3.8 MeV below the Coulomb barrier and then falls after the barrier and reaches towards constant value of 2.1 MeV at higher energies. 

Similarly, in the case of $^{14}$N+$^{90}$Zr, the magnitude of imaginary part of the potential rises as incident energy increases till the peak of the Coulomb barrier and then gets saturated at higher energies. The real part also rises from 2.46 MeV to 4.0 MeV near the Coulomb barrier and then falls from the top of the barrier to a constant value of 2.2 MeV. The bell-shape in real part and the monotonic change in imaginary part of the potential reflected in theoretical calculations are shown by dotted-lines. Such kind of energy dependence in potential tells about the threshold anomaly in the elastic scatterings of both the systems.

\subsection{Essentiality of Non-trivial Behavior}
\begin{figure}
    \centering
    \begin{minipage}{0.45\textwidth}
        \centering
        \includegraphics[width=0.9\textwidth]{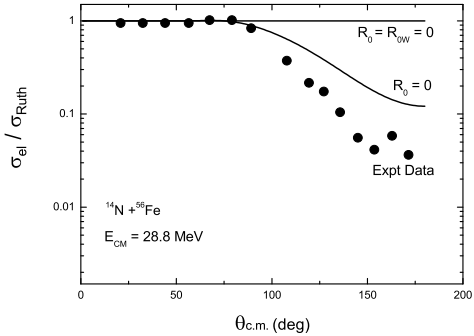} 
        \caption{The comparison of angular elastic scattering cross-section 
	    with experimentally measured data at $E_{CM}$=28.8 MeV with 
	    $R_0$ =$R_{0w}$=0, $R_0$=0 and experimental result for 
	    $^{14}$N+$^{56}$Fe.}
        \label{fig7}
    \end{minipage}
\end{figure}

Boztsun et al.\cite{boztsun2002} have considered additional terms $U_1(r)$ and $U_2(r)$ to the real part of the optical potential to explain the oscillatory behaviour in scattering cross sections of $^{16}$O+$^{28}$Si and $^{12}$C+$^{24}$Mg at the higher angles. These additional terms are simply the derivatives of the Woods-Saxon shape. Additional terms modify the shape of the potential at the surface region. However, we are having a potential that has the inbuilt non-trivial structure to account for such microscopic processes in HIC. This non-trivial feature distinguishes the shape of the volume and surface region. Due to the presence of non-trivial feature in our potential, the same oscillatory behaviour is explained without such addition to the real part\cite{mallick2006}. While trying to fit the elastic scattering data without invoking the non-trivial feature in our potential, we fail to reproduce the experimental observation. The non-trviality in real part is removed if we take $R_0$=0. Here we show the results with $R_0$ = 0 in Fig.\ref{fig7} which doesn't explain the data. Similarly, by removing the non-trviality in imaginary part i.e., $R_{0W}$=0 along with $R_0$=0, we show the results in Fig.\ref{fig7} for the system  $^{14}$N+$^{56}$Fe at energy $E_{CM}$=28.8 MeV. This non trivial nature help reduce the depth of imaginary part of the potentials.
\subsection{Variation of Reflection Function}
\begin{figure}
    \centering
    \begin{minipage}{0.45\textwidth}
        \centering
        \includegraphics[width=0.9\textwidth]{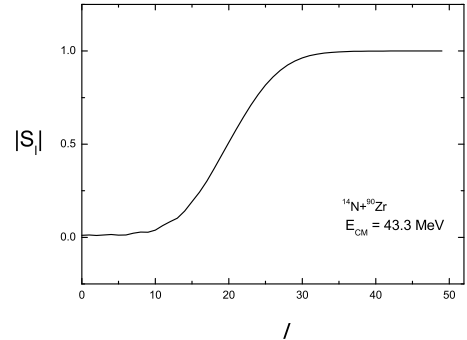} 
        \caption{Plot of reflection function $|S_l|$ versus $l$ for the 
	    system $^{14}$N+$^{90}$Zr at $E_{CM}$ =43.3 MeV with all the 
	    parameters of potential as indicated in 
	    Table-\ref{table_paramet_zr}.}
        \label{fig8}
    \end{minipage}
\end{figure}

\begin{figure}
    \centering
    \begin{minipage}{0.45\textwidth}
        \centering
        \includegraphics[width=0.9\textwidth]{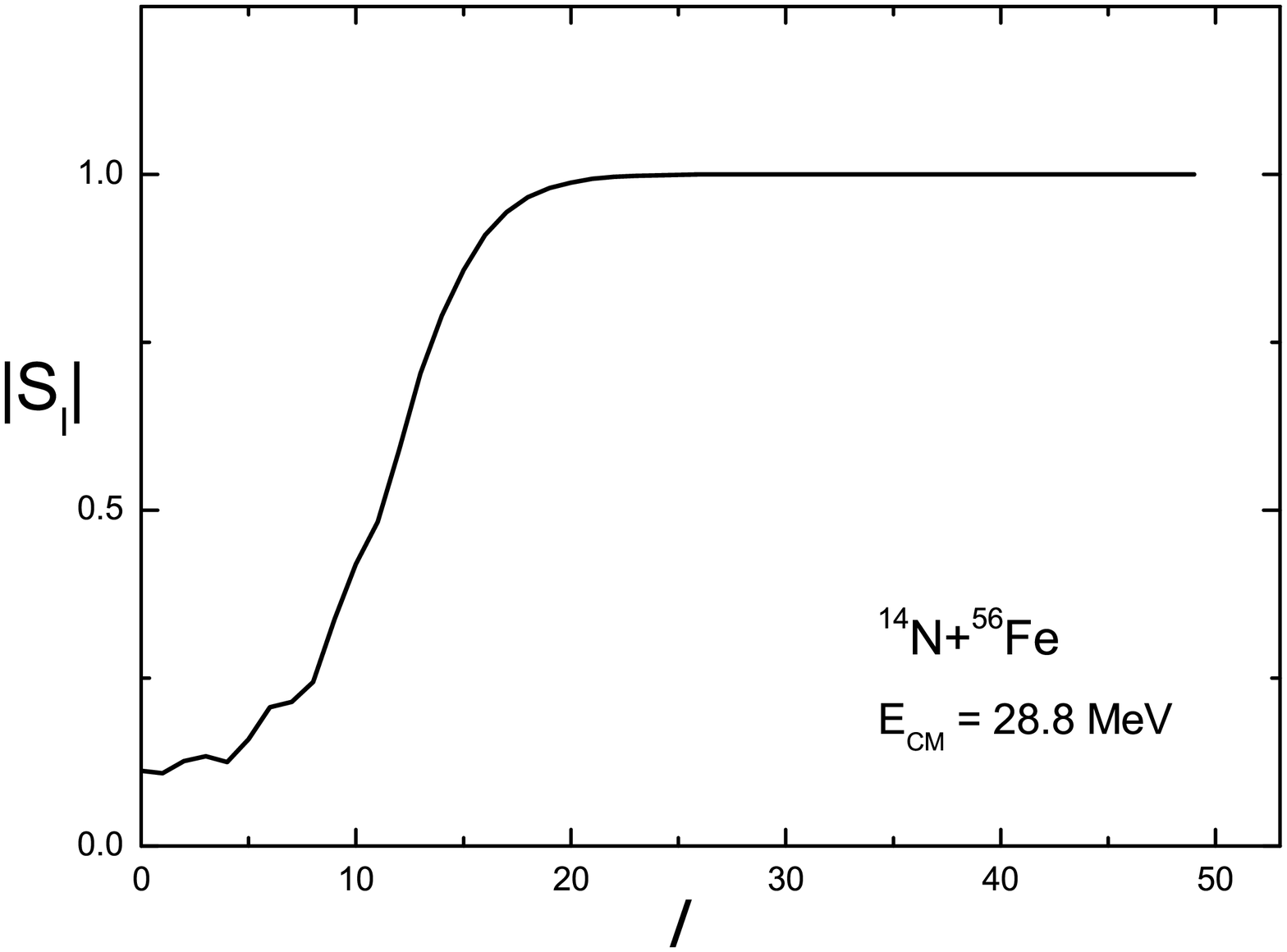} 
        \caption{Plot of reflection function $|S_l|$ versus $l$ for the system $^{14}$N+$^{56}$Fe at $E_{CM}$ = 28.8 MeV}
        \label{fig11}
    \end{minipage}
\end{figure}

From Eq.\ref{eq_sigmael} we know that the behaviors of the magnitude and phase of S-matrix are important for elastic scattering. Knowing the importance of the partial wave S-matrix $S_l$ that determines the scattering amplitude and crosssection, we discuss $|S_l|$ for both the systems. In Fig.\ref{fig8}, the reflection function $|S_l|$ for the system $^{14}$N+$^{90}$Zr is plotted for various ‘$l$’  at incident energy $E_{CM}$= 43.3 MeV. The variation is shown for $l$=0 to $l=$50. This figure tells about the contribution of partial waves. Low partial waves are absorbed to a greater extent and contribute towards nuclear reactions while the partial waves with ‘$l$’ greater than twelve are effective in the Coulombic region. 
The absorptive domain is clearly distinguished from pure coulombic domain with large transition region. That means $\Delta l$ is having a large value. It is already proposed in \cite{santosh2005jpg} that a large value of $\Delta l$ which indicates the dominance of more number of partial waves in the transition region, leads to a smooth behaviour in the $\sigma_{el}/\sigma_{Ruth}$ in the heavy projectile system. A similar behaviour is also observed here in the case of $^{14}$N+$^{90}$Zr (Fig.\ref{fig8}) and $^{14}$N+$^{56}$Fe (Fig.\ref{fig11}).

It is worth to mention that a large number of partial
waves enter the interior region of the potential because of
small value of $V_{BW}$ at the surface region. Barrier waves
got scattered from the barrier and internal waves entered
the interior region. The superposition of barrier waves and internal waves in the presence of a non-trivial feature in the effective potential produces the desired elastic scattering results. The phenomenon like transfer of nucleons, effects of frictional forces, shape deformation of interacting nuclei etc. during heavy-ion collision can affect the nature of phenomenological optical potential.
\section{\label{sec_conclude} CONCLUSIONS}
In summary, we use a phenomenological optical potential with small imaginary part which is developed from Ginocchio potential. The potential exhibits threshold anomaly and explain the data of differential cross section ratios of elastic to Rutherford, $\sigma_{el}/\sigma_{Ruth}$ for systems, $^{14} N+ ^{56} Fe$ and $^{14} N+ ^{90} Zr$ at various colliding energies. This potential has significantly less parameters. By using this potential authors in\cite{mallick2006} were able to explain differential scattering cross-sections of $^{16}$O+$^{28}$Si and $^{12}$C+$^{24}$Mg over a wide range of energies. Present calculation with very small imaginary potential, compared to the one taken in \cite{williams75}, explain the experimental observation. 

The potential considered here has a specific deformation effect at the surface of a nucleus. This deformation controls the absorption during nucleus-nucleus scattering. This is manifested in the present analysis for systems $^{14}$N+$^{56}$Fe and $^{14}$N+$^{90}$Zr and for systems, $^{16}$O+$^{28}$Si and $^{12}$C+$^{24}$ Mg in \cite{mallick2006}. 
So it may be inferred that the shape of the potential at the surface of 
the nucleus may cause deformation near Coulomb barrier. This potential exhibit threshold anomaly near the Coulomb barrier for the systems $^{14}$N+$^{56}$Fe and $^{14}$N+$^{90}$Zr. For various projectile, including weakly bound systems, the analysis with present potential will be reported soon. \\
{\bf Acknowledgement}: Authors thank Tapan Kumar Rana for helpful comments and critical reading of the manuscript.
\twocolumngrid


\begin{thebibliography}{99}
\bibitem{ginocchio84}J. N. Ginocchio, Ann. Phys. (N.Y.) 152,203 (1984).
\bibitem{hodgson86}P. E. Hodgson,  Rep. Prog. Phys., 34, 765 (1971).
\bibitem{mahaux86}C. Mahaux et al., Nucl. Phys. A 449, 354 (1986).
\bibitem{kobos84}A. M. Kobos, G. R. Satchler, Nucl. Phys. A 427, 589 (1984).
\bibitem{nagarajan85}M.A. Nagarajan, et al., Phys. Rev. Lett. 54, 1136 (1985).
\bibitem{thompson89}I. J. Thompson et al., Nucl. Phys. A 505, 84 (1989).
\bibitem{stefanini87}A. M. Stefanini et al., Phys. Rev. Lett. 59, 25 (1987).
\bibitem{pereira89}D. Pereira et al., Phys. Lett. B 220, 347 (1989).
\bibitem{diaz89}J. Diaz et al., Nucl. Phys. A 494, 311 (1989).
\bibitem{fulton85}B. R. Fulton et al., Phys. Lett. B 162, 55 (1985).
\bibitem{lin2001}C. J. Lin et al., Phys. Rev. C, 63, 064606 (2001).
\bibitem{abriola89}D. Abriola et al., Phys. Rev. C 39, 546 (1989).
\bibitem{stachler91}G. R. Satchler, Physics Reports, 199, 147 (1991).
\bibitem{byron92}F. W. Byron and R. W. Fuller, Math of Classical and 
	Quantum Physics, Dover Publications,Inc.,New York, 1992, art.6.6, p.340.
\bibitem{hussein2006}M.S. Hussein, P.R.S. Gomes, J. Lubian, L.C. Chamon, Phys. Rev. C 73 (2006) 044610.
\bibitem{figueira2007}J.M. Figueira, J.O.F. Niello, D. Abriola, A. Arazi, O.A. Capurro, E.d. Barbará, G.V.Martí, D.M. Heimann, A.E. Negri, A.J. Pacheco, I. Padrón, P.R.S. Gomes, J. Lubian, T. Correa, B. Paes,Phys. Rev. C 75 (2007) 017602.
\bibitem{pakou2004}A. Pakou, N. Alamanos, G. Doukelis, A. Gillibert, G. Kalyva, M. Kokkoris, S.Kossionides, A. Lagoyannis, A. Musumarra, C. Papachristodoulou, N. Patronis,
G. Perdikakis, D. Pierroutsakou, E.C. Pollacco, K. Rusek, Phys. Rev. C 69 (2004) 054602.
\bibitem{biswas2008}M. Biswas, S. Roy, M. Sinha, M. Pradhan, A. Mukherjee, P. Basu, H. Majumdar, K. Ramachandran, A. Shrivastava, Nucl. Phys. A 802 (1) (2008) 67–81.
\bibitem{souza2007}F.A. Souza, L.A.S. Leal, N. Carlin, M.G. Munhoz, R.L. Neto, M.M.d. Moura, A.A.P. Suaide, E.M. Szanto, A.S.d. Toledo, J. Takahashi, Phys. Rev. C 75 (2007) 044601.
\bibitem{zadro2009}M. Zadro, P. Figuera, A.D. Pietro, F. Amorini, M. Fisichella, O. Goryunov, M. Lattuada, C. Maiolino, A. Musumarra, V. Ostashko, M. Papa, M.G. Pellegriti, F. Rizzo, D. Santonocito, V. Scuderi, D. Torresi, Phys. Rev. C 80 (2009) 064610.
\bibitem{fimiani2012}L. Fimiani, J.M. Figueira, G.V. Martí, J.E. Testoni, A.J. Pacheco, W.H.Z. Cárdenas, A. Arazi, O.A. Capurro, M.A. Cardona, P. Carnelli, E. de Barbará, D. Hojman, D.Martinez Heimann, A.E. Negri, Phys. Rev. C 86 (2012) 044607.
\bibitem{kumawat2008}H. Kumawat, V. Jha, B.J. Roy, V.V. Parkar, S. Santra, V. Kumar, D. Dutta, P.Shukla, L.M. Pant, A.K. Mohanty, R.K. Choudhury, S. Kailas, Phys. Rev. C 78 (2008) 044617.
\bibitem{deshmukh2011}N.N. Deshmukh, S. Mukherjee, D. Patel, N.L. Singh, P.K. Rath, B.K. Nayak, D.C.Biswas, S. Santra, E.T. Mirgule, L.S. Danu, Y.K. Gupta, A. Saxena, R.K. Choudhury,R. Kumar, J. Lubian, C.C. Lopes, E.N. Cardozo, P.R.S. Gomes, Phys. Rev. C 83
(2011) 024607.
\bibitem{santra2011}S. Santra, S. Kailas, K. Ramachandran, V.V. Parkar, V. Jha, B.J. Roy, P. Shukla, Phys. Rev. C 83 (2011) 034616.
\bibitem{shaikh2016}M.M. Shaikh, M. Das, S. Roy, M. Sinha, M. Pradhan, P. Basu, U. Datta, K. Ramachandran, A. Shrivastava, Nucl. Phys. A 953 (2016) 80–94.
\bibitem{palshetkar2014}C.S. Palshetkar, S. Santra, A. Shrivastava, A. Chatterjee, S.K. Pandit, K. Ramachandran, V.V. Parkar, V. Nanal, V. Jha, B.J. Roy, S. Kalias,Phys. Rev. C 89 (2014) 064610.
\bibitem{rodrigo2019}Rodrigo Navarro P\'{e}rez, Jin Lei, Phys. Lett. B, 795,200(2019).





\bibitem{bock65}R. Bock, H. H. Duhm, M. Grosse-Schulte, R. Rdel, 
	Nucl. Phys.A 70, 48 (1965).
\bibitem{rudchik2002}A. T. Rudchik et al., Nucl. Phys. A 700, 2541 (2002).
\bibitem{towsley77}C. W. Towsley et al., Phys. Rev. C 15, 281 (1977).
\bibitem{liu71}M. Liu et al., Nucl. Phys. A 165, 118 (1971).
\bibitem{oertzen70}W. von Oertzen et al., Nucl. Phys. A 143, 34 (1970).
\bibitem{zeller78}A. F. Zeller, G. T. Hickey, D. C. Weissler, D. F. Hebbard, Nucl. Phys. A 301, 130 (1978).
\bibitem{yamaya88}T. Yamaya et al., Phys. Rev. C 37, 2585 (1988).
\bibitem{balster87}G. J. Balster et al., Nucl. Phys. A 468, 93 (1987).
\bibitem{williams75}M. E. Williams et al., Phys. Rev. C 11, 3 (1975).
\bibitem{sahu2003}B. Sahu, G. S. Mallick and S. K. Agarwalla, Nucl. Phys. A 727, 299 (2003).
\bibitem{mallick2006}G.S. Mallick, S. K. Agarwalla, B. Sahu, C. S. Shastry, Phys.Rev. C 73, 054606 (2006).
\bibitem{aygun17} M. Aygun, Chinese J. Phys., 55, 2559 (2017)
\bibitem{bass74}R. Bass, Nucl. Phys. A 231, 45, (1974).
\bibitem{boztsun2002}I. Boztsun et al., Phys. Rev. C 66, 024610 (2002).
\bibitem{santosh2005jpg} S. K. Agarwalla, G. S. Mallick, P. Prema, S. Mahadevan, B. Sahu and C. S. Shastry, J. Phys. G, 32(2),165(2005).

\end{thebibliography}
\end{document}